\documentstyle[12pt,epsf]{article}

\begin{document}
%
\input psfig

\pagestyle{empty}

\title{Hadronic Decays of B Mesons}
\author{ T.E. Browder, University of Hawaii at Manoa}
\date{}
\maketitle

\centerline{University of Hawaii preprint UH-511-863-96} 
\medskip
\centerline{(To appear in the Proceedings of the Third International
Symposium on Radiative Corrections} 
\centerline{in Cracow, Poland and in Acta Polonica).}

\begin{abstract}

We review recent experimental
results on hadronic decays and lifetimes of hadrons containing $b$
and $c$ quarks\cite{bhp}. We discuss charm counting and the
semileptonic branching fraction in B decays, 
the color suppressed amplitude in B decay, 
and the search for gluonic penguins in B decay.

\end{abstract}
\bigskip

\def\Journal#1#2#3#4{{#1} {\bf #2}, #3 (#4)}

\def\NCA{\em Nuovo Cimento}
\def\NIM{\em Nucl. Instrum. Methods}
\def\NIMA{{\em Nucl. Instrum. Methods} A}
\def\NPB{{\em Nucl. Phys.} B}
\def\PLB{{\em Phys. Lett.}  B}
\def\PRL{\em Phys. Rev. Lett.}
\def\PRD{{\em Phys. Rev.} D}
\def\ZPC{{\em Z. Phys.} C}

\def\st{\scriptstyle}
\def\sst{\scriptscriptstyle}
\def\mco{\multicolumn}
\def\epp{\epsilon^{\prime}}
\def\vep{\varepsilon}
\def\ra{\rightarrow}
\def\ppg{\pi^+\pi^-\gamma}
\def\vp{{\bf p}}
\def\ko{K^0}
\def\kb{\bar{K^0}}
\def\al{\alpha}
\def\ab{\bar{\alpha}}
\def\be{\begin{equation}}
\def\ee{\end{equation}}
\def\bea{\begin{eqnarray}}
\def\eea{\end{eqnarray}}
\def\CPbar{\hbox{{\rm CP}\hskip-1.80em{/}}}

\bibliographystyle{unsrt}    

\section{Charm counting and the semileptonic branching fraction}
\subsection{The Experimental Observations}\label{subsec:prod}

A complete picture of inclusive
$B$ decay is beginning to emerge from recent measurements 
by CLEO~II and the LEP experiments.
These measurements can be used to address
the question of whether the hadronic decay of the B meson is compatible with
its semileptonic branching fraction.

Three facts emerge from the experimental examination of inclusive $B$ decay:
$$n_c = 1.15 \pm 0.05$$ where $n_c$ is the number of charm quarks
produced per $B$ decay taking an average of ARGUS, CLEO 1.5, and CLEO II
results and using  ${\cal B}(D^0\to K^-\pi^+)=(3.76\pm 0.15\%)$\cite{jdr}.

$${\cal B}(B\to X\ell\nu)=10.23\pm 0.39 \% $$ This value is the 
average of the CLEO and ARGUS model independent measurements using dileptons.
The third quantity is calculated from the inclusive $B\to D_s$, 
$B\to (c\bar{c}) X$, and $B\to \Xi_c$ branching fractions,
$${\cal B}(b\to c \bar{c} s)=0.158\pm 0.028\% .$$ This
value is determined assuming
no contribution from $B\to D$ decays, 
an assumption which can be checked using data.    

\subsection{Theoretical Interpretation}\label{subsec:wpp}

In the usual parton model, it is 
difficult to accomodate a low semileptonic branching fraction unless the 
hadronic width of the B meson is increased.\cite{bsl}

The explanations for the semileptonic branching fraction
which have been proposed can be formulated by expressing
the hadronic width of the $B$ meson in terms of three components:
$$\Gamma_{hadronic}(b)  = \Gamma (b\to c \bar{c} s) 
+ \Gamma  (b\to c \bar{u} d) +\Gamma (b\to s~g).$$ If the semileptonic
branching fraction is to be reduced to the observed level, then one of these
components must be enhanced.

A large number of explanations have been proposed in the last few years.
These explanations can be logically classified as follows:

\begin{enumerate}
\item An enhancement of $b\to c \bar{c} s$ due to large QCD corrections
or the breakdown of local duality\cite{bccs1},\cite{bccs2},
\cite{bccs3},\cite{bccs4}.

\item An enhancement of $b\to c \bar{u} d$ due to non-perturbative effects
\cite{bcud1},\cite{bcud2},\cite{bcud3},\cite{bcud4}.

\item An enhancement of $b\to s~g$ or $b\to d~g$ 
from New Physics\cite{new1},\cite{new2},\cite{new3}.

\item The cocktail solution: For example, 
if both the $b\to c \bar{c} s$ and the 
$b\to c\bar{u} d$ mechanisms are increased,
this could suffice to explain the inclusive observations.

\item There might also be a 
 systematic experimental problem in the determination of either $n_c$,
${\cal B}(b\to c \bar{c} s)$, or $ {\cal B}(B\to X \ell\nu)$\cite{isisys}.
\end{enumerate}

\subsection{Other experimental clues}

Inclusive charm particle-lepton correlations can be used to probe
the $B$ decay mechanism and give further insight into this problem. 
High momentum leptons are used $p_{\ell}>1.4$ GeV
to tag the flavor of the B. The angular correlation between the meson
and the lepton is then employed to select events in which the tagging lepton
and meson are from different $B$s. When the lepton and meson
originate from the same B meson they tend to be back to back, whereas when
the meson and leptons come
from different B mesons they are uncorrelated. 
After this separation is performed, wrong sign correlations from $B-\bar{B}$ 
mixing must be subtracted. Since the mixing rate is well measured, 
this correction is straightforward and has little uncertainty.

This technique has been applied previously to several types of 
correlations of charmed hadrons and leptons.
For example, the sign of $\Lambda_c$-lepton
correlations distinguishes between the 
$b\to c \bar{u} d$ and the $b\to c \bar{c} s$ mechanisms.
It was found that the $b\to c\bar{c} s$ mechanism comprises
$ 19\pm 13\pm 4 \%$ of $B\to \Lambda_c$ decays\cite{lamlep}. 
Similiarly, examination of $D_s$-lepton correlations shows that
most $D_s$ mesons originate from $b\to c \bar{c} s$ rather than
from $b\to c \bar{u} d$ with $s \bar{s}$ quark popping at the lower vertex.
In this case, it was found that $17.2\pm 7.9\pm 2.6\%$ of the $D_s$ mesons
originate from the latter mechanism\cite{dslep}.
The same experimental 
technique has now been applied to $D$-lepton correlations.

The conventional $b\to c\bar{u} d$ mechanism which was {\it 
previously assumed} to be
responsible for all $D$ production in $B$ decay will give $D-\ell^+$
correlations. 
If a significant fraction of $D$ mesons
arise from $b\to c\bar{c} s$ with light quark popping at the
upper vertex.
This new mechanism proposed by Buchalla, Dunietz,
and Yamamoto will give $D-\ell^-$ correlations\cite{bccs2}.

Preliminary results of this study have been presented by CLEO~II which finds,
$\Gamma(B\to D~X)/\Gamma(B\to \bar{D} X) = 0.107\pm 0.029\pm 0.018$.\cite{kwon}
This implies a new contribution to the $b\to c \bar{c} s$ width
$${\cal B}(B\to D X) = 0.081\pm 0.026.$$ ALEPH finds evidence for 
$B\to D^0\bar{D^0} X + D^0 D^{\mp} X$ decays with a substantial branching 
fraction of $12.8\pm 2.7\pm 2.6 \%$\cite{alephdd}. 
DELPHI reports the observation of
$B\to D^{*+} D^{*-} X$ decays with a branching fraction 
of $1.0\pm 0.2\pm 0.3\%$\cite{delphidd},\cite{feindt}. 
Since CLEO has set upper limits on the
Cabibbo suppressed exclusive decay modes
$B\to D \bar{D}$ and $B\to D^* \bar{D^*}$ in the $10^{-3}$
range,\cite{cleodd} 
this implies that the signals observed by ALEPH and DELPHI involve
the production of a pair of $D$ mesons and additional particles.
The rate observed by ALEPH is consistent with the rate of wrong sign
$D$-lepton correlation reported by CLEO.
It is possible that these channels are actually resonant modes
of the form $B\to D {D}_s^{**} $
decays, where the p-wave $D_s^{**}$ or radially excited $D_s^{'}$ 
decays to $\bar{D}^{(*)} \bar{K}$\cite{blokddk}.

We can now recalculate $${\cal B}(b\to c \bar{c} s) = 0.239\pm 0.038,$$
which would suggest a larger charm yield ($n_c \sim 1.24$).
This supports
hypothesis (1),  large QCD corrections in $b\to c \bar{c} s$. 
{\it BUT the charm yield $n_c$ as computed in the usual way
is  unchanged}. The contribution of 
$B\to D \bar{D} K X$ was properly accounted for
in the computation of $n_c$. This suggests that the experimental
situation is still problematic.
Is there an error in the normalization ${\cal B}(D^0\to K^-\pi^+)$ ?
Or is there still room for enhanced ${\cal B}(b\to c u \bar{d})$ ?

We note that ALEPH and OPAL have recently reported a value for 
$n_c$ in $Z\to b\bar{b}$ decay\cite{alephnc},\cite{opalnc}. ALEPH finds 
$n_c^Z = 1.230\pm 0.036 \pm 0.038 \pm 0.053$.
The rate of $D_s$ and $\Lambda_c$ production is significantly higher
than what is observed at the $\Upsilon(4S)$. It is not clear whether
the quantity being measured is the same as $n_c$ at the $\Upsilon(4S)$, which
would be the case if the spectator model holds and if the contribution
of $B_s$ and $\Lambda_b$ could be neglected. 
OPAL reports a somewhat lower value of $n_c =1.10\pm 0.045\pm 0.060\pm 0.037$
after correcting for unseen charmonium states. 
OPAL assumes no contribution from $\Xi_c$ production while ALEPH includes
a large contribution from this source. The
contribution of $B\to $baryon decays to charm counting 
as well as the $\Lambda_c$, $Xi_c$ branching fraction scales are still poorly
understood and merit further investigation.

There are other implications of these observations.
A $B$ decay mechanism with a ${\cal O}(10\%)$ branching 
fraction has been found
which was not previously included in the CLEO or LEP
Monte Carlo simulations of $B$ decay. This may have consequences
for other analyses of particle-lepton correlations. For example,
CLEO has re-examined the model independent dilepton measurement
of ${\cal B}(B\to X \ell\nu)$. Due to the lepton threshold of 0.6 GeV
and the soft spectrum of leptons, the CLEO measurement
is fortuitously unchanged. It is also important to check the size
of this effect in LEP measurements of the $B$ semileptonic branching fraction
using dileptons.

\section{The sign of the color suppressed amplitude and lifetimes}

The sign
and magnitude of the color suppressed amplitude can be determined using
several classes of decay modes in charm and bottom mesons. The numerical
determination assumes factorization and uses form factors from various
phenemonological models.

For $D$ decay one
uses exclusive modes such as $D\to K\pi$, $D\to K\rho$ etc., 
and obtains 
$$ a_1 = 1.10\pm 0.03,~ a_2 = -0.50\pm 0.03  $$
The destructive interference observed in two body $D^+$
decays leads to the $D^+$-$D^0$ lifetime difference.

For $B$ decay, one can find the magnitude of $|a_1|$ from
the branching fractions for the decay modes
$\bar{B}^0\to D^{(*)+}\pi^-$, $\bar{B}^0\to D^{+(*)}\rho^-$.
This gives $|a_1|=1.06\pm 0.03\pm 0.06$.
One can also extract $|a_1|$
from measurements of branching fractions
$B\to D^{+,(0)} D_s^{(*)-}$. 
The magnitude $|a_2|$ can be determined from the branching
fractions for $B\to \psi K^{(*)}$. This yields $|a_2|=0.23\pm 0.01\pm 0.01$.

\begin{figure}[htb]
\centerline{\epsfysize 3.5 truein\epsfbox{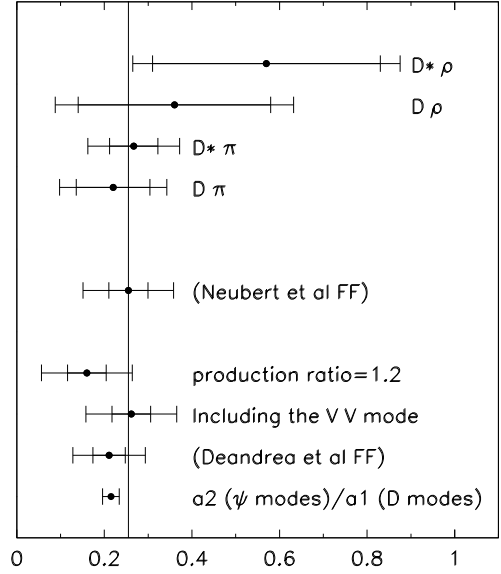}}
\caption{Values of $a_2/a_1$ determined from
beauty decay modes. The top four points show the values extracted
from the individual decay modes.
The bottom four points with error bars
show how the result changes when various assumptions are modified.}
\label{a2a1}
\end{figure}

The value of $a_2/a_1$ can be found by comparing
$B^-$ decays where both the external and spectator diagrams
contribute to $\bar{B}^0$ decays where only the external spectator
decays contribute. For example,
the model of Neubert et al. predicts the following ratios:
\begin{equation}
R_1 = {{\cal B}(B^- \to D^0 \pi^-) \over {\cal B}(\bar{B^0}\to D^+ \pi^-)}
                = (1 + 1.23 a_2/a_1)^2  \label{colrate1}
\end{equation}
\begin{equation}
R_2 = {{\cal B}(B^- \to D^0 \rho^-)
\over {\cal B}(\bar{B^0} \to D^+ \rho^-)}
                = (1 + 0.66 a_2 /a_1)^2  \label{colrate2}
\end{equation}
\begin{equation}
R_3 = {{\cal B}(B^- \to D^{*0} \pi^-)
         \over {\cal B}(\bar{B^0} \to D^{*+} \pi^-)}
                     =(1 + 1.29 a_2/a_1)^2  \label{colrate3}
\end{equation}
\begin{equation}
R_4 = {{\cal B}(B^- \to D^{*0} \rho^-)
          \over{\cal B}(\bar{B^0} \to D^{*+} \rho^-)}
                     \approx (1 + 0.75 a_2/a_1)^2   \label{colrate4}
\end{equation}

Using the latest branching fractions, 
we find $$ a_2/a_1 = 0.26 \pm 0.05 \pm 0.09,$$ where the second
error is due to the uncertainty ($\sim 20\%$)
in the relative production of $B^+$ and
$B^0$ mesons at the $\Upsilon(4S)$. There are a number of uncertainties
which could significantly modify the magnitude of $a_2/a_1$. For example,
the ratios of some heavy-to-heavy to heavy-to-light form factors is needed
(e.g. $B\to \pi/B\to D$). An estimate of this uncertainty
is given by comparing the value of $a_2/a_1$ 
determined using
form factors from the model of Neubert {et al.} with the value obtained
using form factors from the
model of Deandrea {\it et al.}. We also  note the effect
of including the $B\to V V$ mode for which the form factors have somewhat 
larger theoretical uncertainties. 
It is important to remember that
this determination also assumes the factorization hypothesis.
From Fig.~\ref{a2a1} it is clear that
the large error on the relative production of $B^+$ and $B^0$ mesons
is the most significant uncertainty in the determination of $a_2/a_1$. 

\begin{figure}[htb]
\centerline{\epsfysize 3.5 truein\epsfbox{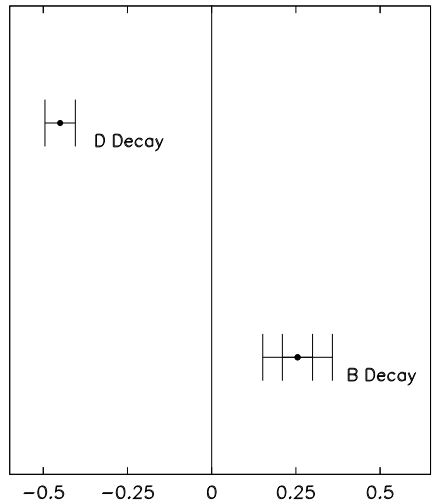}}
\caption{Values of $a_2/a_1$ determined from hadronic charm and
hadronic beauty decay modes}
\label{a2a1_charm}
\end{figure}

As shown in Fig.~\ref{a2a1} this
is consistent with the ratio $|a_2|$/$|a_1|$ where $|a_2|$ is computed
from $B\to \psi$ modes and $|a_1|$ is computed from $\bar{B}^0\to D^{(*)}\pi,
D^{(*)}\rho$ modes. Although the result is surprisingly different
from what is observed in hadronic charm decay (see Fig.~\ref{a2a1_charm})
and from what is expected in the $1/N_c$ expansion,
Buras claims that the result can be accomodated
in NLO QCD calculations\cite{buras}.

If the constructive interference which is observed in these
$B^+$ decays is present in all $B^+$ decays, then we expect
a significant $B^+$-$B^0$ lifetime difference ($\tau_B^{+}< \tau_{B^0}$), 
of order $15-20\%$. This is only marginally consistent 
with experimental measurements of lifetimes;
the world average computed in 
our review\cite{bhp} is $$\tau_{B^+}/\tau_{B^0}= 1.00\pm 0.05 .$$

It is possible 
that the hadronic $B^+$ decays that have been observed to date are 
atypical. The remaining higher multiplicity $B^+$ decays could
have destructive interference or no interference. Or perhaps 
there is a mechanism which also enhances the $\bar{B}^0$
width to compensate for the increase in the $B^+$ width
and which maintains the $B^+/B^0$ lifetime ratio near unity.
Such a mechanism would be relevant to the charm counting and
semileptonic branching fraction problem.
In either case, there will be experimental consequences in the
pattern of hadronic $B$ branching fractions. CLEO can 
experimentally compare other $B^-$ and $B^0$ decays including 
$D^{**}\pi^-$ and $D^{**}\rho^-$ as well
decays to $D^{(*)} a_1^-$, $a_1^-\to \rho^0\pi^-$ 
and $D^{(*)} b_1^-$, $b_1^-\to \omega\pi^-$ to check the first
possibility.

\section{The search for the gluonic penguin}

It is important to measure the size of ${\cal A}(b\to s~g)$,
the amplitude for the gluonic penguin,  in order
to interpret the CP violating asymmetries which will be observed at
future facilities. Gluonic penguin modes will also be used to search
for direct CP violation at future facilities.

\begin{figure}[htb]
\centerline{\epsfysize 3.0 truein\epsfbox{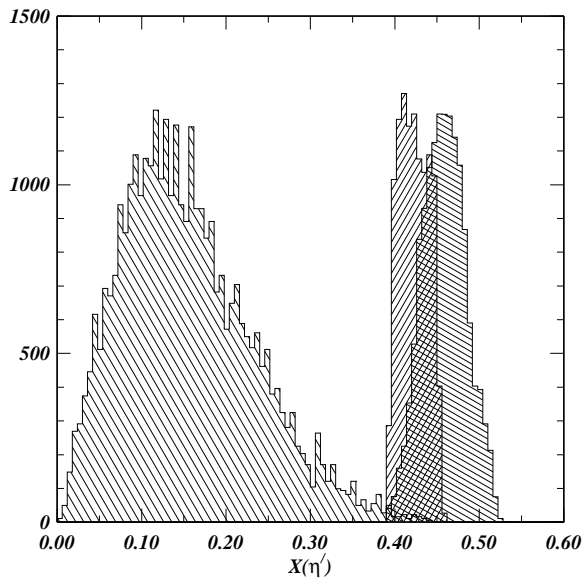}}
\vskip 15mm
\caption{Monte Carlo distributions for inclusive $B\to \eta^{'}$ production
as a function of scaled $\eta^{'}$ momentum. The relative normalization
is arbitrary. The $b\to c$ mechanism is dominant below $x=0.4$.
The region above $x>0.4$ may contain contributions from internal spectator 
decays such as $\bar{B}^0\to D^{(*)} \eta^{'}$ (cross-hatched to the
left) as well as $b\to s g^*$, $g^*\to q\bar{q}$ decays 
such as $B\to K^{(n)} \eta^{'}$ (cross-hatched to the right).}
\label{etaprime_mc}
\end{figure}

CLEO-II has observed a signal in the sum of $\bar{B}^0\to K^+ \pi^-$
and $\bar{B}^0\to \pi^+ \pi^-$ with a branching fraction of
$(1.8^{+0.6+0.2}_{-0.5-0.3})\times 10^{-5}$ and for the individual modes
${\cal B}(B^0\to \pi^+\pi^-)<2.0\times 10^{-5}$,
${\cal B}(B^0\to K^+\pi^-)<1.7\times 10^{-5}$. Similiar results with
consistent branching fractions have been 
reported by DELPHI\cite{delphikpi} and ALEPH\cite{alephkpi}.
CLEO-II has also observed a signal in the sum of $B^-\to K^- \omega$
and $B^-\to \pi^- \omega$.\cite{omegah} 
The combined branching fraction is
$(2.8\pm 1.1\pm 0.5) \times 10^{-5}$. 
In all of these cases, due to the paucity of
events and the difficulty of distinguishing high momentum kaons and pions,
the conclusion is that either $b\to u$ or $b\to s~g$ decays
or a combination of the two processes has been observed.

Another approach using quasi-inclusive decays is described in a recent
CLEO contribution\cite{inclusive}. At the $\Upsilon(4S)$, 
two body decays from $b\to s~g$ can be distinguished
from $b\to c$ decays by examination of the inclusive particle momentum 
spectrum; the $b\to s~g$ decays populate a region beyond the kinematic limit
for $b\to c$. This approach has been applied to inclusive $\eta^{'}$, $K_s$,
and $\phi$ production.

A search for  inclusive signatures of $b\to s$ gluon rather
than exclusive signatures has two possible advantages.
The inclusive rate may be calculable from first principles and is expected to 
be at least an order
of magnitude larger than the rate for any exclusive channel.
For example, the branching fraction for $b\to s q\bar{q}$ 
(where $q=u, ~d,~s$) is
${\cal O}(1\%)$\cite{Grigjanis},\cite{Desh1} and
the branching fraction for the inclusive process
$b\to s\bar{s} s$ is expected 
to be $\sim 0.23\%$ in the Standard Model\cite{Desh1}, 
while low multiplicity decay
modes such as $\bar{B}^0\to \phi K_s$ or $\bar{B}^0\to K^-\pi^+$
are expected to have branching fractions
of order $10^{-5}$.
The disadvantage of employing an inclusive method 
is the severe continuum background that must be
subtracted or suppressed. 

\begin{figure}[htb]
\centerline{\epsfysize 3.0 truein\epsfbox{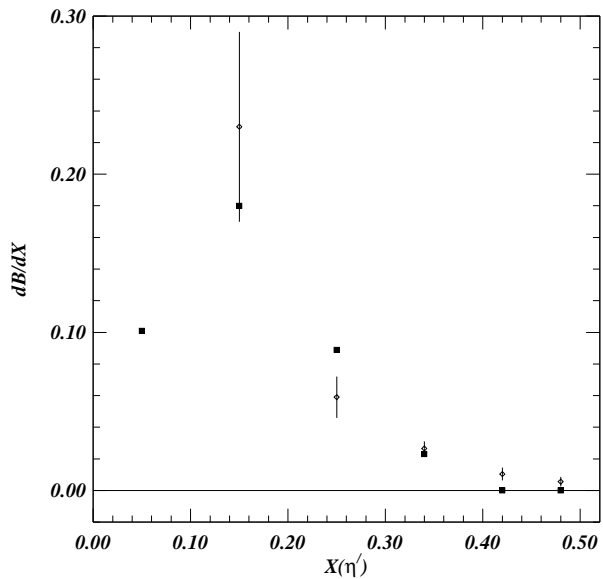}}
\vskip 15mm
\caption{Branching fraction for inclusive $B\to \eta^{'}$ production
as a function of scaled $\eta^{'}$ momentum. The points with error bars
are measurements. The squares are the predictions of the CLEO
Monte Carlo simulation. The kinematic range $0.4<x<0.5$ is used for
the gluonic penguin search.}
\label{etaprime_branch}
\end{figure}

The  decay 
$B \to \eta^{'}  X_s$, where $X_s$ denotes a meson containing
an $s$ quark, is dominated by the gluonic penguins, 
$b\to s g^*$ $g^*\to s \bar{s}$, $g^*\to u \bar{u}$ or
$g^*\to d \bar{d}$. 
The decay $B \to K_s  X$, where $X$ denotes a meson which contains
no $s$ quark, arises from a similiar gluonic penguin, 
$b\to s g^*$ $\to s \bar{d} d$.

An analogous search for $b\to s g^*$, $g^*\to s \bar{s}$
was carried out by CLEO using 
high momentum $\phi$ production\cite{phix}. In the search for
high momentum $\phi$ production, limits were obtained using 
two complementary techniques. A
purely inclusive technique with shape cuts gave a  limit 
${\cal B}(B\to X_s \phi)<2.2 \times 10^{-4}$ for $2.0<p_{\phi}<2.6$
GeV. Using the $B$ reconstruction technique, in which combinations of
the $\phi$ candidate, a kaon, and up to 4 pions were required to be
consistent with a $B$ candidate, gave a limit of 
${\cal B}(B\to X_s \phi)<1.1 \times 10^{-4}$ for $M_{X_s}<2.0$ GeV,
corresponding to $p_{\phi}>2.1$ GeV.
These results can be compared to the
Standard Model calculation of Deshpande {\it et al.}\cite{Desh2},
which predicts that the branching fraction for this process 
should lie in the range $(0.6-2.0) \times 10^{-4}$
and that $90\%$ of the $\phi$ mesons from this mechanism will lie
in the range of the experimental search. 
Ciuchini {\it et al.}\cite{ciuchini} predict a branching fraction for
${\cal B}(B\to X_s \phi)$ in the range $(1.1\pm 0.9)\times 10^{-4}$.
One sees that the sensitivity of the inclusive method is
nearly sufficient to observe a signal from Standard Model $b\to s~g$.

Using the purely inclusive technique, 
a modest excess was observed in the signal region for 
quasi two-body $B\to \eta^{'} X_s$ decays. 
A 90\% confidence level 
 upper limit of for the momentum interval $0.39<x_{\eta^{'}}<0.52$,
$${\cal B}(B\to \eta^{'} X_s) < 1.7 \times 10^{-3}$$
is obtained. Further work is in progress to improve the sensitivity
in this channel.
Examination of high momentum $K_s$ production shows no excess and gives
a 90\% confidence level upper limit of 
$${\cal B}(B\to K_s X) < 7.5
\times 10^{-4}$$ for $0.4 < x_{K_s} < 0.54$. 
We expect that gluonic penguin decays will be observed in the
near future by CLEO.
More theoretical work is required to convert the limits 
presented here as well as future observations into
constraints on the quark level process $b\to s~g^*, g^*\to q \bar{q}$.


\begin{thebibliography}{99}
\bibitem{jdr} J. D. Richman, contribution to the Proceedings of the
ICHEP 96 Conference in Warsaw. The value of $n_c$ is quite sensitive
to the $D$ branching fraction scale. A lower value for $n_c$ is computed by 
Richman by using a slightly higher world average
for ${\cal B}(D^0\to K^-\pi^+)$ that incorporates a recent measurement
from ALEPH.
\end{thebibliography}
\end{document}